\input epsf
\documentstyle[twocolumn,aps,prb]{revtex}
\begin{document}
\draft
\tightenlines
\twocolumn[\hsize\textwidth\columnwidth\hsize\csname@twocolumnfalse\endcsname
\title{The Josephson critical current in a long mesoscopic S-N-S junction}
\author{P.~Dubos, H.~Courtois, B.~Pannetier}
\address{Centre de Recherches sur les Tr\`es Basses
Temp\'eratures-C.N.R.S. associated to Universit\'e Joseph Fourier, 25
Ave. des Martyrs, 38042 Grenoble, France}
\author{F. K.~Wilhelm$,^{1,2}$ A. D.~ Zaikin$^{3}$ and G.~Sch\" on$^{1}$}
\address{$^1$Institut f\" ur Theoretische Festk\" orperphysik, Universit\" at
Karlsruhe, 76128 Karlsruhe, Germany\\
$^2$Quantum Transport Group, Department of Applied Physics and DIMES,
TU Delft, 2600 GA Delft, The Netherlands\\
$^3$Forschungszentrum Karlsruhe, Institut f\"ur Nanotechnologie, 
76021 Karlsruhe, Germany}
\date{\today}
\maketitle
\begin{abstract}
We carry out an extensive experimental and theoretical study of the
Josephson effect in S-N-S junctions made of a diffusive normal metal
(N) embedded between two superconducting electrodes (S).  Our
experiments are performed on Nb-Cu-Nb junctions with
highly-transparent interfaces.  We give the predictions of the
quasiclassical theory in various regimes on a precise and quantitative
level.  We describe the crossover between the
short and the long junction regimes and provide the temperature
dependence of the critical current using dimensionless units
$eR_{\rm N}I_{\rm c}/\epsilon_{\rm c}$ and $k_{\rm B}T/\epsilon_{\rm c}$ where
$\epsilon_{\rm c}$ is the Thouless energy.  Experimental and 
theoretical results are in excellent
quantitative agreement.
\end{abstract}
 ]

The Josephson effect is well known to exist in weak links connecting
two superconducting electrodes S, \makebox{e.g.} a tunnel barrier
I, a short constriction C or a normal metal N (S-I-S, S-C-S and S-N-S
junctions).  This effect manifests itself in a non-dissipative
DC-current flowing through the Josephson junction at zero voltage.  At
weak coupling, e.g. in the S-I-S case, the Josephson current can be
expressed as $I_{\rm s}= I_{\rm c} \sin\varphi$ where $\varphi$ is the
phase difference between the two superconducting condensates and the
maximum supercurrent $I_{\rm c}$ is called the critical current.

The Josephson effect in S-N-S junctions has been studied in a variety
of configurations.  The early experiments of Clarke\cite{Clarke} and
Shepherd\cite{Shepherd} were performed in Pb-Cu-Pb sandwiches.  In
these experiments and in the pioneering calculations by de
Gennes,\cite{deGennes} it was already realized that the presence of a
supercurrent in such structures is due to the proximity effect.
It can be understood as the generation of superconducting correlations
in a normal metal connected to a superconductor, mediated by
phase-coherent Andreev reflections at the S-N interface.
The critical current $I_{\rm c}$ is limited by the ``bottleneck''
in the center of the N-layer, where the pair amplitude is
exponentially small, $I_{\rm c}\propto e^{-L/L_{\rm T}}$.  Here
$L_{\rm T} =\sqrt{\hbar D/2 \pi k_{\rm B} T}$ is the characteristic
thermal length in the diffusive limit and $L$ is the length of the
junction. These calculations, as well as those by Fink,\cite{Fink76} 
analyzed the
temperature dependence of $I_{\rm c}$ within the Ginzburg-Landau theory in
the vicinity of the superconducting critical temperature $T_{\rm c}$.
Later, the
critical current of diffusive S-N-S microbridges\cite{Warlaumont,Dover}
was successfully described by Likharev \cite{Likharev} with
the aid of the quasiclassical Usadel equations.\cite{Usadel} In this
work, the emphasis was put on the high temperature regime where the
superconducting order parameter is smaller than the thermal energy
$\Delta \ll k_{\rm B}T$. A more general
study of the Josephson effect in diffusive S-N-S junctions was made
in Ref. \onlinecite{Zaikin81}.

More recently, experimental data on long Josephson junctions
\cite{Courtois95} showed a surprising temperature-dependence, which
turned out to be in a strong disagreement with the early theory by de
Gennes.  These data have been discussed by some of us
\cite{WilhelmJLTP} within the quasiclassical approach which we will
also use in the present work.  Fink\cite{Fink97} attempted to analyze
the data \cite{Courtois95} by means of an extrapolation of the
Ginzburg-Landau theory to low temperatures.

The proximity effect in mesoscopic hybrid structures consisting of normal
and superconducting metals attracted a growing interest during the recent
years.\cite{Superlattice}
Here we will consider mesoscopic diffusive S-N-S junctions where the sample
length is much larger than the elastic mean free path $l_{e}$ but
smaller than the dephasing length $L_{\rm \phi}$~:
$l_{\rm e}<L<L_{\phi}$. In N-S junctions and Andreev interferometers, we
can identify -- both theoretically and experimentally -- the natural
energy scale for the proximity effect.\cite{CourtoisPrl,Revue_JLTP} It
is given by the Thouless energy $\epsilon_{\rm c}=\hbar D/L^2$.
Here $D=v_{\rm F}l_{\rm e}/3$ is the diffusion constant of the N-metal,
$v_{\rm F}$ is the Fermi velocity.  In contrast to the energy
gap $\Delta$ which is set by the interactions in the superconducting
electrodes, the energy scale $\epsilon_{\rm c}$ is a single-electron
quantity~: $\epsilon_{\rm c}/\hbar$ is merely the diffusion rate across
the sample for a single electron. This energy scale remains important
in non-equilibrium situations, e.g. if one drives the
supercurrent across a S-N-S junction by the injection of a control
current in the N-metal.\cite{vanWees,Transistor_SNS,Lindelof}

The main purpose of the present paper is to carry out a detailed
experimental investigation of the equilibrium supercurrent in
relatively long diffusive S-N-S junctions with highly transparent N-S
interfaces as well as a quantitative comparison of our data to
the theoretical predictions.  Here, a long junction means that the
junction length $L$ is much bigger than $\sqrt{\hbar D/\Delta}$. This 
is equivalent to
$\Delta \gg \epsilon_{\rm c}$. In order to perform this comparison at all
relevant temperatures, we complete the previous studies by
providing a rigorous expression for the Josephson critical current at
$T \to 0$ which was not properly evaluated before. Our experimental
results are in excellent agreement with the theoretical predictions.

As before,\cite{Zaikin81,WilhelmJLTP} our theoretical approach is
based on the quasiclassical Green's functions in imaginary time.  
The proximity effect is described by a finite pair
amplitude $F$ in the N-metal (see [\onlinecite{SuperlatticeFW}] and
references therein).
We will assume N-S interfaces to be fully
transparent and neglect the suppression of the pair potential $\Delta$
in the S electrodes near the N-S interface.  This is appropriate at
$T\ll T_{\rm c}$ or if the reservoirs are very massive as compared to
the normal metal. Within those bounds, our calculation does not contain
further approximations and is e.g. valid at arbitrary temperature and
sample size. We will now proceed by discussing certain limits.

In the high temperature regime $k_{\rm B}T \gg \epsilon_{\rm c}$ (or,
equivalently, $L \gg L_{\rm T}$), the solution is well known.  In this
case the mutual influence of the two superconducting electrodes can be
neglected and the Usadel equations can be linearized in the N-metal,
except in the vicinity of the N-S interfaces.  One finds~:\cite{Zaikin81}
\begin{equation}
e R_{\rm N}I_{\rm c} =\\
64 \pi k_{\rm B}T \sum_{n=0}^{\infty}\frac{L}{L_{\rm \omega_{\rm n}}}
\frac{\Delta^2 \exp(-{L/L_{\rm \omega_{\rm n}}})}{(\omega_{\rm n}
+\Omega_{\rm n}+\sqrt{2(\Omega^2 + \omega_{\rm n} \Omega_{\rm n})})^2}
\label{Matsu}
\end{equation}
where $R_{\rm N}$ is the N-metal resistance, $\omega_{\rm n}=(2n+1)\pi
k_{\rm B}T$ is the Matsubara frequency, $\Omega_{\rm n}=\sqrt{\Delta^2
+ \omega_{\rm n}^2}$ and $L_{\rm \omega_{\rm n}}=\sqrt{\hbar
D/2 \omega_{\rm n}}$.
If $T$ is close to the critical temperature of S, the gap is small as
compared to the thermal energy : $\Delta \ll k_{\rm B}T$.  In this
limit, Eq. (\ref{Matsu}) coincides with the result derived by
Likharev.\cite{Likharev}

\begin{figure}
\epsfxsize=8.5 cm
\epsfbox{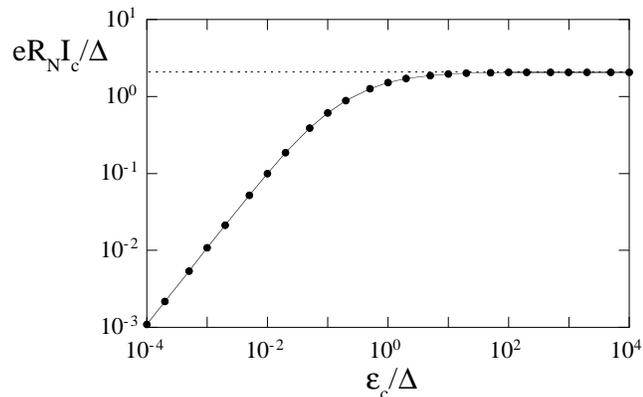}
\caption{Calculated dependence of
the zero temperature $eR_{\rm N}I_{\rm c}$ product in units of
$\Delta$ as a function of the ratio
$\epsilon_{\rm c}/\Delta$.  $I_{\rm c}$ is the Josephson critical
current, $R_{\rm N}$ the normal state resistance, $\epsilon_{\rm c}$ is
the Thouless energy and $\Delta$ is the
superconducting gap of S. The long junction regime is on the left part
of the graph where $\epsilon_{\rm c} < \Delta$, the short junction
regime is on the right part where $\epsilon_{\rm c} > \Delta$.  The
dashed line corresponds to the Kulik-Omel'yanchuk formula\cite{KO} at $T=0$.}
\label{gap/thouless}
\end{figure}

At lower temperatures $k_{\rm B}T \lesssim
\epsilon_{\rm c}$ evaluation of $I_c$ involves solutions of the Usadel
equation at {\em all} energies \cite{SuperlatticeFW}.  In
order to determine the precise value \cite{FN} of the critical current, we
performed a numerical solution of the Usadel equations for the whole
range of Matsubara frequencies. In the long junction
limit ($\Delta \gg \epsilon_{c}$), the zero-temperature $e R_{\rm N}I_{\rm
c}$ is found to be proportional to $\epsilon_{\rm c}$~:
\begin{equation}
e R_{\rm N}I_{\rm c}(T=0)=10.82\epsilon_{\rm c}.
\label{I0}
\end{equation}
In this case, the current phase relation is slightly different from a sine
and the supercurrent maximum occurs at $\varphi=1.27 \pi/2$.%\cite{These_Dubos}
As compared
to previous estimates,\cite{WilhelmJLTP,Fink97} the exact numerical
prefactor in this formula turns out to be unexpectedly high. This 
observation is
crucial for a quantitative comparison between theory and experiment not
only in the case of conventional junctions but also for
high-$T_{\rm c}$ S-N-S junctions \cite{Delin} or devices involving
carbon nanotubes.\cite{Kazu}

Let us now turn to the short junction regime $\Delta \ll \epsilon_{\rm c}$, i.e. to the 
case of dirty S-C-S weak
links described in Ref. \onlinecite{KO,ZP85}. Our numerical
results reproduces quantitatively the behaviours of both the current-phase
relation and the zero-temperature critical current : $eR_{\rm N}I_{\rm c}
\approx 1.326\pi\Delta/2$ at $\varphi = 1.25 \pi/2$.\cite{KO,ZP85} This
results confirms the precision of our calculation in describing 
both the long junction and the short junction regimes.
Our numerical results for $I_{\rm c}(T=0)$ as a function of the
Thouless energy $\epsilon_{\rm c}$ are presented in Fig.
\ref{gap/thouless}. It confirms that it is the minimum
of the gap $\Delta$ and the Thouless energy $\epsilon_{\rm c}$ which
limits the critical current in diffusive S-N-S junctions. At
$\epsilon_{\rm c} \simeq \Delta$, the critical current value remains 
close to the
short junction case.

\begin{figure}
\epsfxsize=8.5 cm \epsfbox{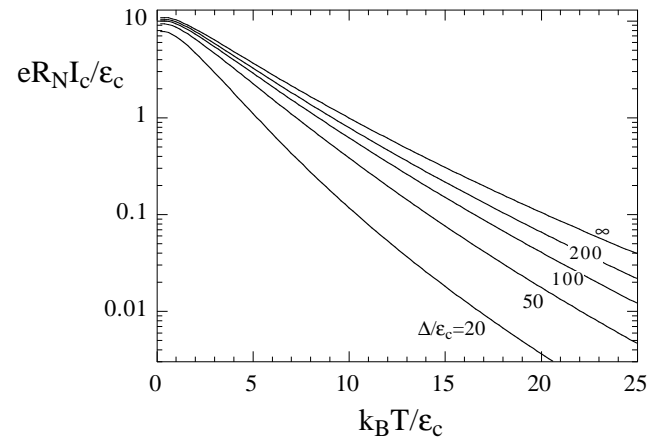} \caption{Calculated temperature
dependence of the $eR_{\rm N}I_{\rm c}$ product.  The different curves
correspond to various values of the ratio $\Delta / \epsilon_{\rm c}$
in the long juction regime.
The curve for $\Delta / \epsilon_{\rm c} \rightarrow \infty $ is
universal in the sense it does not depend on $\Delta$.  Note that
$k_{\rm B}T/\epsilon_{\rm c} = L^{2}/2\pi L_{\rm T}^{2}$.}
\label{IcT_theorique}
\end{figure}

In the following, we will focus on long junctions $\Delta >
\epsilon_{\rm c}$. Fig. \ref{IcT_theorique} shows the temperature 
dependence of the $e
R_{\rm N} I_{\rm c}$ product for various values of the superconducting
gap in the long junction regime.  Both axis are
given in units of the Thouless energy.  The low-temperature part
($k_{\rm B}T < 5 \epsilon_{\rm c}$) comes from a numerical solution of
the Usadel equation while the high-temperature part comes from Eq.
\ref{Matsu}.  From this figure, we can see that the characteristic
decay temperature for the critical current is a few times the Thouless
temperature $\epsilon_{\rm c}/k_{\rm B}$.  As soon as $k_{\rm B}T >
5\epsilon_{\rm c}$, the sum in Eq.  (\ref{Matsu}) can be reduced to
the first frequency term within a $3\%$ underestimation.  This
term corresponds to $\omega_{\rm 0} =\pi k_{\rm B}T$ and
$L_{\rm \omega_{\rm 0}}=L_{\rm T}$.  Adding the second term in the
summation decreases the error below $0.1\%$ in the same temperature
range.

The universal curve of Fig. \ref{IcT_theorique} for $\Delta/\epsilon_{\rm
c} \rightarrow \infty$ is valid only in the case of a very long
junction with $\Delta/\epsilon_{\rm c} \gg 100$.  It appears as if
$\Delta$ is to be compared to the quantity $eR_{\rm N}I_{\rm
c}(T=0)\simeq 10 \epsilon_{\rm c}$
in the long junction limit. In the limit of infinite
$\Delta/\epsilon_{\rm c}$, Eq.  \ref{Matsu} simplifies to
\begin{equation}
e R_{\rm N}I_{\rm c} =\frac{32}{3+2\sqrt{2}}\, \epsilon_{\rm c} \,
\left[\frac{L}{L_{\rm T}}\right]^3 e^{-L/L_{\rm T}}.
\label{Loi_en_T}
\end{equation}
 From Eq. \ref{Loi_en_T}, one can get the
temperature dependence of the critical current : $I_{\rm c} \propto
T^{3/2} exp(-L/L_{\rm T})$.  It has been demonstrated in Ref.
[\onlinecite{WilhelmJLTP}] that within a limited temperature interval
this expression is {\it numerically} very close to a simple
exponential dependence $I_{\rm c} \propto\exp(-L/L_{\rm T})$ with
$L_{\rm T} \propto 1/T$, as one would expect in a ballistic
limit.\cite{Kulik,Ishii} From Fig.  \ref{IcT_theorique}, the
quasi-exponential temperature dependence of the critical current is
indeed striking.  This was the central result of Ref.
[\onlinecite{Courtois95}], but was not understood at that time.  This
coincidence is purely accidental and has no special meaning.\cite{WilhelmJLTP}
In the low temperature limit, the numerical solution can be
approximated by  $e R_{\rm N} I_{\rm c}/\epsilon_{\rm c}=a
(1-b e^{-a \epsilon_{\rm c}/3.2 k_{\rm B} T})$.
The coefficients $a$ and
$b$ are $10.82$ and $1.30$ respectively in the long junction limit
$\Delta/\epsilon_{\rm c}  \rightarrow \infty$.

S-N-S junctions are intrinsically shunted and have negligible internal
capacitance, so they are strongly overdamped. Their current-voltage 
characteristics are hence 
intrinsically non-hysteretic. The transition from a supercurrent to a voltage 
state happens at the critical current, but is rounded by finite
temperature.\cite{AmHalp} We fabricated Nb-Cu-Nb junctions\cite{LT22} 
with a large conductance so
that thermal fluctuations remain small compared to the Josephson 
energy : $k_{B}T \ll \hbar I_{c}(T)/2e$ even at high temperature near the
critical temperature of Nb. This makes a well-defined critical current
up to the critical temperature of Nb. Effects 
of environmental fluctuations known from mesoscopic tunnel
junctions,\cite{Vion} which are intrinsically underdamped, are absent.

\begin{figure}
\center \epsfxsize=5.5 cm \epsfbox{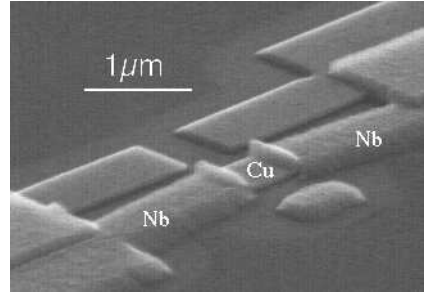} \vspace{0.5 cm}
\caption{Oblique micrograph of a typical S-N-S junction made of a
Cu wire embedded between two Nb electrodes.  The doubling of every
structure due to the shadow evaporation is visible.  The Nb
electrodes cover the Cu strip over about $150 \, nm$.}
\label{Photo}
\end{figure}

We benefited from a trilayer
stencil mask technology\cite{Victrex} making use of a thermostable
resist that does not outgas during Nb evaporation. Thus we were able
to routinely obtain a superconducting critical temperature as high as
$8.1\, K$ for the Nb electrodes.  We performed successive shadow
evaporations of Cu and Nb at different angles through the silicon
stencil layer in an ultra high vacuum chamber, followed by a lift-off.
Fig. \ref{Photo} shows a typical sample.
We studied a single sample (a) plus five different samples evaporated
on the same substrate (b, c, d, e and f).  Table 1 lists the main
physical parameters for these samples.  The Cu metallic strips are
$600 \, nm$ wide and $100 \, nm$ thick.  The Nb superconducting
electrodes are $800 \, nm$ wide and $200 \, nm$ thick, except for
sample a where it is $100 \, nm$.  The length $L$ of the metallic
island was varied between $700$ and $1000 \, nm$, corresponding to a
separation length $d_{\rm Nb}$ between Nb electrodes varying between
$370$ and $700 \, nm$.  For all samples, the calculated Thouless
energy $\hbar D/L^2$ is therefore significantly smaller than the gap
$\Delta$.

The normal-state resistance $R_{\rm N}$ cannot be directly measured at
temperature above $T_{\rm c}$ since the resistance of the Nb
electrodes is measured in series.  We found that the finite-bias
resistance ($eV \simeq \epsilon_{\rm c}$) varied by about $10\%$
between $2\, K$ and $8 \,K$ due to the proximity effect on the
conductance.  We took for the normal-state resistance $R_{N}$ the
resistance at $T = 6\, K$ for a better agreement with the theory.  It
is a relatively high temperature since $k_{\rm B}T > 15 \epsilon_{\rm
c}$ for every sample.  Using L for the Cu length, we obtain a Cu
resistivity $\rho = 1.1\,.\,10^{-8} \,\Omega.m$ for samples b to f and
$\rho = 1.5\,.\,10^{-8} \,\Omega.m$ for sample a.

\begin{figure}
\epsfxsize=8.5 cm \epsfbox{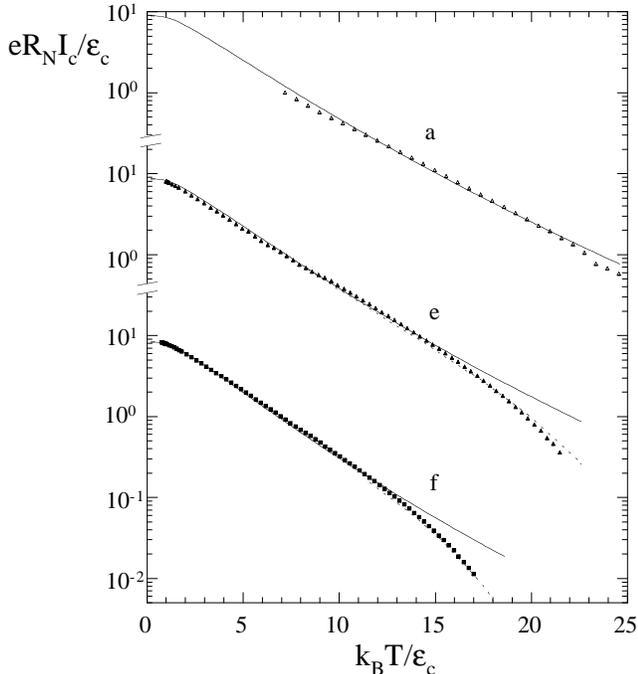}
\caption{Temperature dependence of the measured $eR_{\rm N}I_{\rm
c}$ product of samples a, e and f together with the theoretical fits
assuming a temperature-independent gap (full line) and a gap
following a BCS temperature dependence with $T_{\rm c}=7.5\, K$
(dashed line).  The only adjustment parameter is the Thouless
energy $\epsilon_{\rm c}$ of each sample.  For description of the
sample parameters, see Table 1.}
\label{tous}
\end{figure}

We measured the critical current of samples a to f at temperatures
down to $300 \, mK$. Our procedure consists in sweeping the bias
current while measuring the differential resistance $dV/dI$.
We define the experimental critical current as the current where the
differential resistance reaches $R_{\rm N}/2$. With this criteria, the
experimental uncertainty is estimated below $0.5\%$ at $T=2\,
K$, $5\%$ around $T=4\, K$ and $100\%$ at $7\, K$.
Fig.  \ref{tous} shows the data for 3 samples. The measured $e
R_{\rm N}I_{\rm c}/\epsilon_{\rm c}$ plotted as a function of the
reduced temperature $k_{\rm B}T/\epsilon_{\rm c}$ show a large
decrease over more than two decades.  For each sample, we fitted the
data to the theoretical prediction with only one fitting parameter,
the Thouless energy.  The zero-temperature superconducting gap
$\Delta$ was calculated from the measured critical temperature of Nb
using : $\Delta = 3.8\, k_{\rm B}T_{\rm c}$.\cite{Ashcroft} This gives
$1.3 \, meV$ for all samples except sample a for which $\Delta = 1 \,
meV$.  We used both a fixed gap equal
to the zero-temperature value and a gap $\Delta(T)$ following the BCS
temperature dependence, but with a slightly reduced critical
temperature $T_{\rm c}=7.5\, K$.  At high temperature, it appeared 
necessary to take into account
the temperature dependence of the gap. In this case, the agreement 
between theory and
experiment is excellent.  The fit is very sensitive to the chosen
value of the Thouless energy.  We would like to stress that for each
sample the horizontal and vertical axis are normalized to {\it the
same} Thouless energy $\epsilon_{\rm c}$.  This fitted values are
found to be very close to the Thouless energies calculated from the
full length $L$ of the Cu strip, see Table \ref{samples}.

In Fig.  \ref{tous}, the critical current of sample f shows the onset
of the saturation regime.  At $T\, =\, 300mK$ the adjusted critical
current $e R_{\rm N} I_{\rm c}$ reaches up to $8.2 \, \epsilon_{\rm
c}$.  This number is close to the theoretical value
$8.79\epsilon_{\rm c}$ for sample f at $T=0$.  This result
discards an interpretation of our data within the Ginzburg-Landau
theory of Ref.  \onlinecite{Fink97} which predicts a maximum $eR_{\rm
N}I_{\rm c}/\epsilon_{\rm c}$ of about $1$.

In Ref.  \onlinecite{Courtois95}, an array of S-N-S junctions was made
of a long N-metal wire periodically in contact with a series of
superconducting islands.  A good fit between the data and the theory
was shown in Ref.  [\onlinecite{WilhelmJLTP}], but with the
introduction of a strong reduction of the effective area.  This may be
attributed to the periodic and lateral characters of this type of
samples.

Our calculation assumes perfectly transmitting interfaces with zero
boundary resistance.  In fact, it is sufficient that the
barrier-equivalent length \cite{Spivak} $L_{\rm t}=l_{\rm e}/t$ is
much smaller than the sample length.  As an example, this condition
means an interface transparency $t >
0.1$ for sample b.  In the case of Nb-Cu-Nb samples fabricated through a
two-lithography-step process including Ar-etching,\cite{Mailly} we
found a critical current with a reduced magnitude, presumably due to
a slightly degraded interface. The critical
current in S-N-S junctions with partially transparent interfaces was
discussed in Ref. [\onlinecite{Kuprianov}].  The predicted
behavior features a different temperature dependence for the critical
current. Nevertheless, the temperature dependence remained consistent
with the theory assuming a perfect interface. Only a reduction prefactor
had to be introduced.  This observation could hint at the fact, that
interface barriers are very inhomogenous and the current is carried
through a few highly conducting pinholes.

In summary, we discussed the Josephson critical current of diffusive
S-N-S junctions.  This study provides a simple and reliable
formulation that enables the practical determination of the
equilibrium critical current. We studied the
critical current of a set of samples with different junction lengths
and found an excellent agreement between our data and the predictions of
the quasiclassical theory.

We acknowledge discussion and financial support in the EU-TMR network
"Dynamics of superconducting circuits" as well as support from the DFG
through SFB 195 and GK 284.  We thank A. Golubov, T.T.\ Heikkil\"a, D.
Mailly, N. Moussy and P. Paniez for discussions.

\onecolumn
\begin{center}
\begin{table}
\begin{tabular}{cccccccccc}
\hspace{.9cm} & \hspace {.5cm} & \hspace {.5cm} & \hspace {.5cm} &
\hspace {.5cm}& \hspace {.5cm} & \hspace {.5cm} & \hspace {.5cm}\\
$\#$  & L & $d_{\rm Nb}$ & w  & $R_{\rm N,6K}$ & D  & $\hbar D/L^2 $ &
$\epsilon_{\rm c}$ & $\Delta/\epsilon_{\rm c}$ &
$\frac{eR_{N}I_{c}}{\varepsilon_{c}}(T=0)$\\

        &(nm)  & (nm)& (nm)& $(\Omega)$ & $(cm^2/s)$ &
$(\mu eV)$ & $(\mu eV)$ &  & \\
\hline
a & 1000 & 600 & 600 & 0.260  & 200 & 13 & 14.3 & 70 & 8.91 \\ %D071
b & 1010 & 680 & 580 & 0.173  & 300 & 20 & 18.6 & 70 & 8.99 \\ %A1
c & 910  & 570 & 590 & 0.179  & 260 & 22 & 21.7 & 60 & 8.83 \\ %A2
d & 800  & 470 & 580 & 0.183  & 230 & 25 & 25.4 & 51 & 8.64 \\ %A4
e & 800  & 476 & 590 & 0.169  & 250 & 26 & 26.1 & 50 & 8.62 \\ %B4
f & 710  & 370 & 580 & 0.152  & 250 & 34 & 33.5 & 39 & 8.32 \\ %A5
\end{tabular}
\vspace{0.5 cm}
\caption{Parameters of the different samples studied. $L$ is the full
length of Cu strip while $d_{\rm Nb}$ is the Nb electrodes separation.
$w$ is the Cu strip width. The Thouless energy $\epsilon_{\rm c}$ is
derived from the fit of the experimental data to the theoretical
prediction (see Fig. \ref{tous}).}
\label{samples}
\end{table}
\end{center}


\begin{references}
\bibitem{Zaikin81} A. D. Zaikin and G. F. Zharkov, {\it Sov. J. Low Temp.
Phys.} {\bf 7}, 184 (1981).  In Eq. 14 and 15 of that paper, the
prefactor 128 refers to the case where $R_N$ is the resistance {\em per spin}.
It should be replaced by 64 if one identifies $R_N$ with the
full resistance.
\end{references}
\end{document}